\documentclass{kluwer}
\usepackage[dvips]{graphicx} 
\begin{document}
\begin{article}
\begin{opening}
\title{The Nature of Submm-Detected Galaxies}            
%\subtitle{Basic Instructions}

\author{D. \surname{de Mello}}
\author{T. \surname{Wiklind}} 
\institute{Onsala Space Observatory, SE 43992 Sweden}                               
\author{C. \surname{Leitherer}}
\institute{STScI, Baltimore MD21218 USA}
\author{K. \surname{Pontoppidan}}
\institute{Leiden Observatory, NL-2300 RA Leiden, The Netherlands}
\end{opening}

%\date: rather not

%\dedication{To Jim}

%\translation{De Kluwer LaTeX stylefile; aanwijzingen voor auteurs}

\runningtitle{Submm-Detected Galaxies}
\runningauthor{T. Wiklind}

%\begin{abstract} 
%This document describes how to format a paper with
%\texttt{kluwer.cls}, the Kluwer
%classfile for journal submissions.
%\end{abstract}


%\classification{JEL codes}{D24, L60, 047}
%\end{opening}
\section{Introduction}

The submm detected sources at high redshifts are believed to comprise a class of objects which are highly
luminous dusty sources, emitting most of their luminosity at FIR
wavelengths (see review by I. Smail in these proceedings).
It is presently unclear what is the most common dominative power source: star formation or AGN activity?
In the case of star formation, the implied star formation rates exceed 1000
M$_{\odot}$/yr. Combined with a large gas reservoir, this SFR can be sustained
for relatively long time periods ($\sim$ 10$^{8}$ years) and
the end result is a massive galaxy, created at high redshift in a short time period.
If AGN activity is the power source, however, the observed redshifted FIR and molecular
gas may not directly probe the star formation properties.
In order to study two of the identified submm sources in more detail we have observed
SMM J14011+0252 at z=2.565 and SMM J02399-0136 at z=2.808 with the VLT/FORS2 (longslit), 
with an observed spectral resolution of 1.06 \AA. This is 15 times better than anything achieved previously. The
goal was to search for diagnostic stellar absorption lines, giving information about
the massive stellar population. 

%\section{Results}

\begin{figure}
\centerline{\includegraphics[width=20pc]{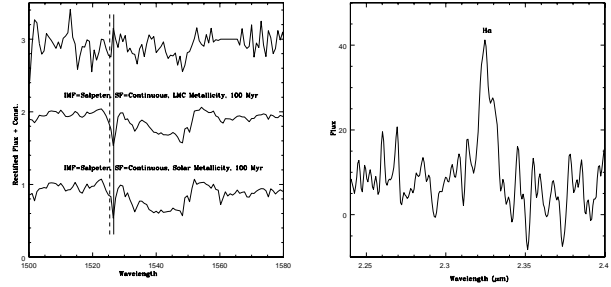}}
\caption[]{Left: VLT/FORS2 spectrum of SMM J14011+0252 (top) and STARBURST99 models (IMF=Salpeter, 
Star Formation Law=continuous, metallicity=LMC (middle) and Solar (bottom), age=100Myr). 
The vertical line is
centered on the interstellar line SiII 1526.7. The vertical dashed line is blue shifted by
1.46 \AA. 
Right:VLT/ISAAC spectrum of SMM J14011+0252. H$\alpha$ is
marked.}
\end{figure}

\begin{figure}
\centerline{\includegraphics[width=20pc]{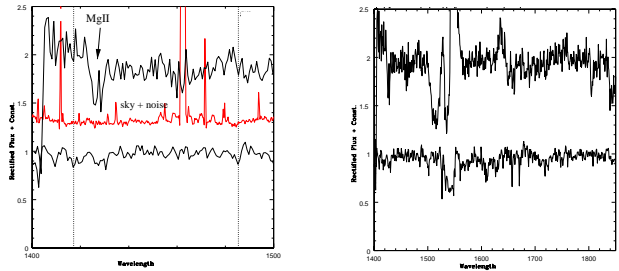}}
\caption[]{Top: VLT/FORS2 spectrum of SMM J02399-0136 and Bottom: the STARBURST99 model for IMF=Salpeter, 
Star Formation Law=continuous, metallicity=solar, age=100 Myr. The intervening MgII 2798 is
marked by an arrow. SiIII 1417 and SiII 1485 are marked by vertical lines. 
The spectrum in the middle of the first panel corresponds to the sky plus the noise level 
in the spectrum.}
\end{figure}

The high resolution optical spectra, probing 1400 - 2000 \AA\ in the UV restframe, show
several stellar absorption lines. For instance, the 
{\bf SMM J14011+0252} spectrum is typical of starbursts. We have compared the data 
with our models of star-forming galaxies {\footnote{STARBURST99 is available at {\tt http://www.stsci.edu/science/starburst99}}}. Fig. 1 (left panel) shows the spectrum of SMM J14011+0252 together with
STARBURST99 model with the following input parameters: Star Formation Law = continuous, IMF = Salpeter and 
metallicity = LMC and solar, age=100 Myr. The dashed vertical line is centered on the interstellar
line SiII 1526.7 which is blue shifted in SMM J14011+0252 (outflow velocity $\sim$ 290 km/s). The shape of CIV 1550
suggests that this object has metallicity similar to the LMC. The H$\alpha$ profile in the 
VLT-ISAAC spectrum shows no sign of an AGN. On the other hand, the {\bf SMM J02399-0136} spectrum 
shows clear signs of an AGN (CIV~1550 in emission) as previously known. However, we have also found 
evidences of a starburst, a host galaxy as well as an intervening galaxy at z=0.94. These results are based on (i) the absorption lines which are 
starburst diagnostics marked by vertical lines on Fig. 2 (SiIII~1417 and SiII~1485); (ii) the morphology of the galaxy
seen in our VLT images; and (iii) the absorption line identified as MgII~2798 absorption lines at z=0.94. The presence of
 an intervening galaxy is of great importance regarding the understanding of this peculiar hyperluminous
 galaxy. It might mean an extra magnification factor which is probably what make this
object such a hyperluminous galaxy identified as a SCUBA source.
We are reducing more data, which will improve the signal to noise of our spectra.

\end{article}
\end{document}